\documentclass[apj,twocolumn]{emulateapj}

\slugcomment{accepted in ApJ}

\shortauthors{Parmentier \& Fritze}
\shorttitle{Cluster Formation Efficiency}

\begin{document}


\title{When efficient star formation drives cluster formation}


\author{G.~Parmentier\altaffilmark{1,2,3} \& U.~Fritze\altaffilmark{4}}


\altaffiltext{1}{Belgian Science Policy Research Fellow and Humboldt Fellow}
\altaffiltext{2}{Institute of Astrophysics \& Geophysics, University of Li\`ege, 
All\'ee du 6 Ao\^ut 17, B-4000 Li\`ege, Belgium}
\altaffiltext{3}{Argelander-Institut f\"ur Astronomie, University of Bonn,
Auf dem H\"ugel 71, D-53121 Bonn, Germany}
\altaffiltext{4}{Centre for Astrophysics Research, University of Hertfordshire, 
Hatfield AL10 9AB, United Kingdom}


\begin{abstract}

We investigate the impact of the star formation efficiency in cluster-forming-cores
(i.e. local SFE) on the evolution of the mass in star clusters over the age 
range 1-100\,Myr, when star clusters undergo their infant weight-loss/mortality phase.  
Our model builds on the $N$-body simulation grid of \citet{bau07}.  
Assuming a constant formation rate of gas-embedded clusters and a weak tidal field,
we show that the ratio between the total mass in stars bound to the clusters over 
that age range and the total mass in stars initially formed in gas-embedded clusters 
is a strongly increasing function of the averaged local SFE, with little influence 
from any assumed core mass-radius relation.  Our results suggest that, for young starbursts 
with estimated tidal field strength and known recent star formation history, observed 
cluster-to-star mass ratios, once corrected for the undetected clusters, constitute promising 
probes of the local SFE, without the need of resorting to gas mass estimates.

Similarly, the mass ratio of stars which remain in bound clusters at the end of the infant 
mortality/weight-loss phase (i.e. age $\gtrsim 50$\,Myr) depends sensitively on the mean local
SFE, although the impacts of the width of the SFE distribution function and of the core mass-radius relation
require more careful assessment in this case.  Following the recent finding by \citet{bas08} that 
galaxies form, on the average, 8\,\% of their stars in bound clusters 
regardless of their star formation rate, we raise the hypothesis that star formation in the 
present-day Universe is characterized by a 
near-universal distribution for the local SFE.  A related potential application of our model consists in 
tracing the evolution of the local SFE over cosmological lookback times by comparing 
the age distribution of the total mass in star clusters to that in field stars in galaxies 
where field stars can be resolved and age-dated.  We describe model aspects which
are still to be worked out before achieving this goal.

\end{abstract}


\keywords{galaxies: star clusters --- galaxies: evolution --- galaxies: starburst  --- stars: formation --- stellar dynamics}

\section{Introduction}

Star formation efficiencies (SFE) are not easy to come by observationally since they require reliable determinations of both the gaseous and stellar masses of a star forming region.  Yet, the SFE is a fundamental parameter of star and star cluster formation processes and of the early evolution of star clusters.  {\it Local} SFEs, i.e. the mass fraction of a giant molecular cloud core turned into stars, have been estimated for a handful of Galactic nearby gas-embedded clusters~\citep{ll03}.  They range from 10 to 30 per cents, which is an order of magnitude higher than what is inferred when the SFE is averaged over an entire giant molecular cloud (the so-called {\it global} SFE).  Attempts to compare SFE estimates in the Milky Way and its neighbours, where they are best studied, to those in major merger-induced starbursts, involves the additional difficulty of scale differences.  While for the former, SFEs are defined down to individual giant molecular clouds (GMC) and their dense star forming cores, only large-scale -- often galaxy-wide -- SFEs are measured for the latter owing to their large distances.  
Estimates of SFEs over up to kpc size nuclear regions in Ultra Luminous IR Galaxies (=ULIRGs), all of which are late stages of massive gas-rich mergers, are reported to be $1-2$ orders of magnitude higher than  global SFEs measured in spiral galaxies~\citep[][their fig.~2b]{gao04}.

Because the formation of compact and massive star clusters is observed to be closely associated to active star forming environments, prominent star formation episodes in the history of galaxies are best traced by the age distribution of their star clusters~\citep{West+04,Fritze04,BrodieStrader06,Kotulla+08}.
Although the vast majority of stars actually form in gas-embedded star clusters, galaxies are predominantly made of field stars.  In this respect, a peak in a star cluster age distribution only tells us {\em when} an epoch of enhanced star cluster formation took place, not yet {\em how much total stellar mass was involved}.  To derive the total mass in stars formed in a starburst older than $\sim 3-4$ Gyr from the integrated spectrum of a galaxy is, in the present state-of-the-art, hopeless~\citep{FritzeLilly07}.
Knowing the mass fraction of all star formation still in bound clusters as a function of star cluster age, by properly accounting for cluster destruction and mass loss, would allow us to carry the analysis of galaxies'evolution one step further and to estimate the history of their overall star formation rate.
We now have a fairly mature view of how to decipher the rates of {\it bound} cluster formation and evolution of a given star cluster system, in essence based on the cluster age and mass distributions~\citep{par08a}.  That is, we are able to assess the amount of field stars arising from the secular evolution 
(i.e. the preferential depletion of low-mass stars due to tidal stripping and internal two-body relaxation) of star clusters
which survived the expulsion of their residual star forming gas.  As pointed out by \citet{mcl99}, however, secular evolution does not necessarily affect the total mass in star clusters significantly.  As for the star clusters of the Galactic Old Halo and of the Large Magellanic Cloud for instance, the observed total mass in clusters is less than an order of magnitude smaller than their total mass at the onset of secular evolution~\citep[see table 3 and table 2 in][respectively]{par05, par08a}.  If the vast majority of stars form in gas-embedded star clusters, this implies that the bulk of field stars is given off at an earlier stage, very likely as a result of cluster dynamical response to the expulsion of the residual star forming gas out of their parent cores due to massive star activity.  Because of its resulting shallower potential, an exposed cluster attempts to reach a new equilibrium, which results into the loss of stars (``infant weight-loss``) and, when the local SFE is too low, the eventual disruption of the cluster (``infant mortality``)~\citep[see Fig.~1 in ][]{goo06, goo08}.  This violent relaxation phase, which lasts for at most 50\,Myr, appears therefore as the prime driver of the ratio of the total mass in bound clusters to the total mass in stars.  In what follows, we assume that all stars form in gas-embedded clusters, i.e. we assume that there is no mode of field star formation or loose stellar association formation.  We will further comment about this hypthesis in our discussion (section 3).

Young starbursts are the best places to infer the impact of cluster violent relaxation on the cluster-to-star mass ratio.  Based on {\sl HST} $UV$ imaging of a sample of starburst galaxies, the morphology of which ranges from blue compact dwarfs to ultra-luminous merging far-infrared galaxies, 
\citet{meu95} find that, on average, about 20 per cents of the $UV$ luminosity 
comes from star clusters.  \citet{deg03} study the Mice and Tadpole interacting systems
by means of a pixel-by-pixel analysis of their color-magnitude and color-color diagrams.
In the Mice interacting galaxies, they find that star clusters are predominantly associated 
with regions of active star formation, where they contribute $\geq 40$\,per cents of the total
flux, irrespective of wavelength (but see our discussion section for caveats regarding these values).
Similar flux ratios are inferred for the Tadpole system, 
an example of a galaxy encounter between two unequal-mass galaxies.
These high luminosity fractions, presumably representative of large mass
fractions, demonstrate that star clusters contribute significantly 
to the total stellar light given off by young starbursts.  While being higher than the
initial Galactic halo globular cluster mass fraction as compared to halo field stars, 
i.e. $\simeq 0.1$, these ratios are also conspicuously smaller than unity, thus confirming 
that the dissolution of star clusters takes place on a short time-scale.

Analyses of star cluster systems on the basis of multi-band imaging are feasible out to Virgo cluster distances and have the potential to provide a comprehensive view of the individual star formation, chemical enrichment and mass assembly histories of their parent galaxies over the past Hubble-Time, provided that the last missing issue, establishing the ratio of star formation ending up in bound star clusters, is solved. This is what we attempt here, that is, we investigate what is/are the driving parameter(s) behind the mass fraction of clusters as compared to field stars.  Specifically, we build on the $N$-body simulation grids 
recently obtained by \citet{bau07} to compute the evolution over the age range 1-100\,Myr 
of the mass in clusters and to compare it with the overall star formation 
rate arising from the same star formation episode.  We make predictions about the cluster
formation efficiency (i.e. the ratio between the total mass in clusters over the age
range of relevance and the total mass in star forming gas cores) and the ratio
between the total mass in clusters and the total mass in stars.  And we highlight their relation to the local SFE.

\section{Method and Results}
\label{sec:res}

The starting point of our simulations is an ensemble of molecular cores, the high-density regions of GMCs associated to star formation.  The core population is statistically described in the following way.   The core mass spectrum follows a power-law ${\rm d}N \propto m^\alpha {\rm d}m$ of spectral index $\alpha =-2$ and of lower and upper mass limits $m_l=100\,M_{\odot}$ and $m_{up}=10^7\,M_{\odot}$, respectively.  
Each cluster forming out of one of these cores is assigned an age $t$ randomly drawn from a uniform age distribution ranging between 1 and 100\,Myr.  Cluster ages will be used to compute their mass and to build the temporal evolution of the mass in star clusters through the violent relaxation phase.
Each of our Monte-Carlo simulations encompasses $10^6$ cores with mass $m_c$, radius $r_c$ (the core radius is defined below) and corresponding cluster age $t$.  Combined to the assumed core mass spectrum, this core number is equivalent to a total mass of star forming gas of $1.1 \times 10^9\,M_{\odot}$.  We note that the core number has been chosen so as to limit Poissonian noise in the gas-embedded cluster formation rate we will derive, rather than as a match to a particular galaxy.  We account for star formation with a time-independent Gaussian probability distribution function of the SFE of mean $\bar\epsilon$ and standard deviation $\sigma _\epsilon$, which we denote  $P(\epsilon)=G(\bar\epsilon, \sigma _\epsilon)$.  We emphasize that $\epsilon$ is the {\it local} SFE, that is, the core mass fraction turned into stars, as opposed to global SFE measured on galactic-wide or GMC-wide scales.  The corresponding constant star formation rate is thus $\bar\epsilon \times 11\,M_{\odot}.yr^{-1}$, equivalent here to the total stellar mass formed per unit of time in gas-embedded clusters.

To obtain the stellar mass $m_{\rm cl}(t)$ still bound to a cluster of age $t$ through the infant mortality/infant weight-loss phase, the gas-embedded cluster mass $\epsilon \times m_c$ is multiplied by the bound fraction $F_b$ of stars at age $t$:
\begin{equation} 
m_{\rm cl}(t) = F_{\rm b}(t) ~\epsilon ~ m_{\rm c}\;.
\label{eq:minit}
\end{equation}
Stellar groups of which the stellar mass is less than $100\,M_\odot$ are not considered as star clusters and their mass is not accounted for in the total cluster mass.  Because of their small total mass, this is of negligible impact on the cluster-to-star mass ratios we derive.  
In the $N$-body grid of \citet{bau07}, from which our star cluster system simulations are derived, $F_b(t)$ is defined as the stellar mass fraction of the initially gas-embedded cluster which resides within the instantaneous cluster tidal radius.  It depends on the local SFE $\epsilon$, on the age $t$ of the cluster, on the gas removal time-scale $\tau _{\rm GR}/\tau _{\rm cross}$ (expressed in unit of a protocluster crossing-time) and on the strength of the external tidal field.  The older the cluster and/or the lower the star formation efficiency and/or the quicker the gas expulsion and/or the stronger the host galaxy tidal field, the lower the bound fraction $F_b$.
To recover this four-fold dependence of $F_b$, we interpolate the output files of the $N$-body simulation runs carried out by \citet{bau07}.  Their grid of results provides the evolution with time of $F_b$ following gas expulsion, as a function of $\epsilon$, of $\tau _{\rm GR}/\tau _{\rm cross}$ and of the ratio  of the half-mass radius to the tidal radius of the gas-embedded cluster $r_h/r_t$, which quantifies the strength of the tidal field.

\begin{figure}
\begin{center}
\vspace*{-3mm}
\epsscale{1.2} \plotone{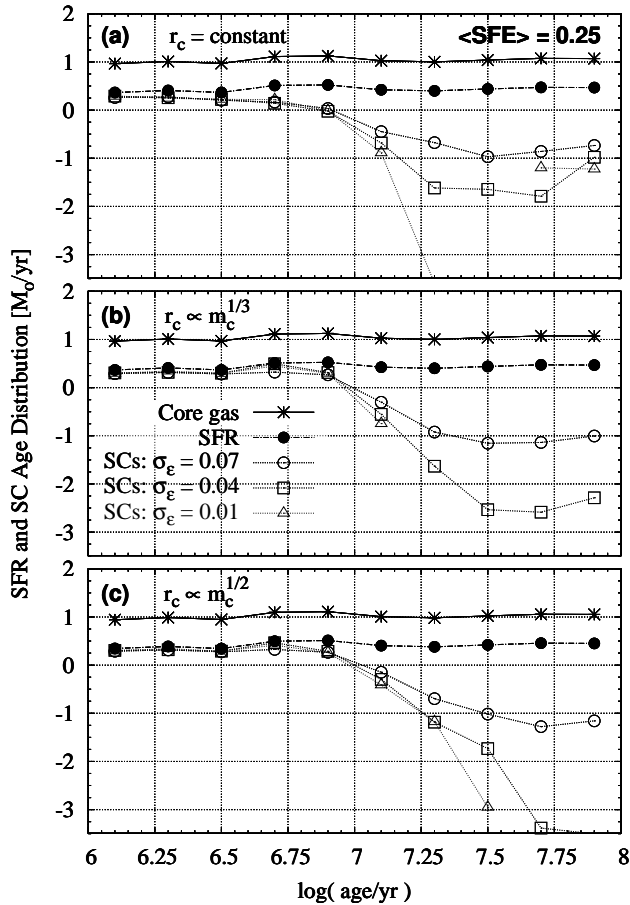}
\caption{Time evolution of the mass in clusters over the course of violent relaxation.  The mass of core gas available to cluster formation per time unit (solid lines with asterisks) and the distribution function of the star formation efficiency are assumed constant in time, resulting into a constant star formation rate (dash-dotted lines with plain circles).  The age distribution of clusters in terms of mass counts is illustrated for different core mass-radius relations (panels a, b and c) and for different standart deviations of the star formation efficiency probability distribution (dotted lines with open circles, squares and triangles).  The mean star formation efficiency is $\bar\epsilon =0.25$}
\label{fig:cfr0.25}
\end{center} 
\end{figure}

\begin{figure}
\begin{center}
\epsscale{1.2} \plotone{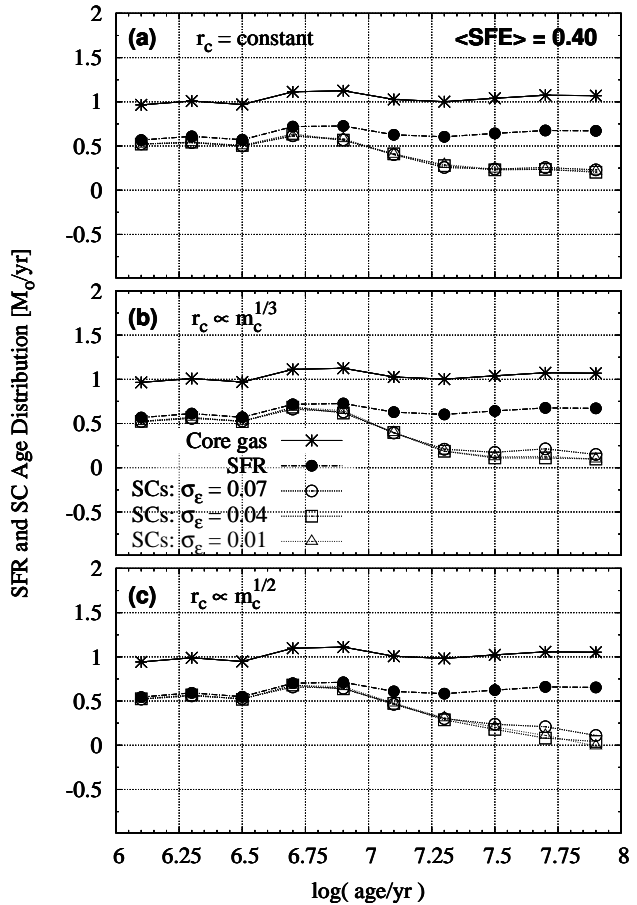}
\caption{Same as in Fig.~\ref{fig:cfr0.25}, for a mean star formation efficiency of $\bar\epsilon =0.40$.  Note the difference in the y-scaling with  Fig.~\ref{fig:cfr0.25}}
\label{fig:cfr0.40}
\end{center} 
\end{figure}

\begin{figure}
\begin{center}
\epsscale{1.2} \plotone{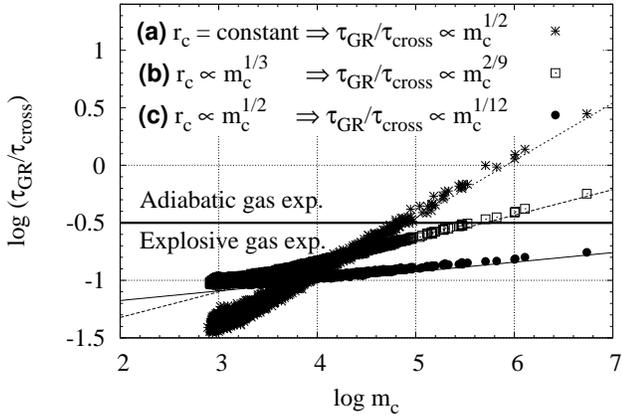}
\caption{Relations between the gas expulsion time-scale in unit of one protocluster crossing time and the cluster forming core mass, for different core mass-radius relations.  The steeper the core-mass radius relation, the weaker the dependence of the gas expulsion time-scale on the core mass.  A sample only of the one million cores simulated is shown}
\label{fig:tt_mc}
\end{center} 
\end{figure}

The local SFE and the age $t$ are sampled from their respective distribution functions.
The gas removal time-scale $\tau _{\rm GR}/\tau _{\rm cross}$ is defined by equation 6 in \citet{par08b}: it depends on the local SFE $\epsilon$ and on the core mass $m_c$ and radius $r_c$.  It remains unclear how $r_c$ scales with $m_c$.  Theoretical  expectations are that the mass-radius relation of virialized cores obeys $r_c \propto m_c^{1/2}$~\citep[e.g.][]{hp94}.  However, this is at variance with observations that the radii of gas-embedded clusters and recently exposed clusters do not vary significantly ($r \simeq 1$\,pc) over several orders of magnitude of cluster masses~\citep[see Table 1 in][]{kro05}.
Considering how debated this issue remains, in what follows, we consider three distinct cases: cores with constant radius ($r_c \propto m_c^0$), constant density ($r_c \propto m_c^{1/3}$) and constant surface density ($r_c \propto m_c^{1/2}$, i.e. virialized cores).  Corresponding simulation results are respectively shown as cases {\it (a)}, {\it (b)} and {\it (c)} in Figs.\ref{fig:cfr0.25}, \ref{fig:cfr0.40} and \ref{fig:tt_mc}, and in Table \ref{tab:clusterfrac}.  As for the normalization of these relations, we refer to Figure 1 of \citet{tan07}, which shows the mass, radius, volumic mass density and surface mass density of gas-embedded Galactic protoclusters.  In case {\it (a)}, we assume that the protocluster radius is a good proxy to the core radius and we adopt $r_c=0.7$\,pc.  In case {\it (b)}, we fit by eye a constant volumic mass density line to the Galactic protoclusters and we multiply the corresponding density by about a factor of 3, so as to account for the mass contribution of the residual star forming gas (i.e. we assume a mean local SFE of 0.33). This gives: $r_c=0.026 \times (m_c/1M_{\odot})^{1/3}$\,pc.  In case {\it (c)}, adopting a similar approach for a constant surface mass density, we find $r_c=0.008 \times (m_c/1M_{\odot})^{1/2}$\,pc.  Lower normalizations (i.e. smaller radius for a given mass) would slow down gas expulsion, thereby increasing the fraction of stars remaining bound to the cluster~\citep[see equation 6 and Figure 1 in][]{par08b}.  In order to highlight the impact of gas expulsion only on the cluster age distribution, we limit our simulations to the case of a weak tidal field.  The vast majority of our gas-embedded clusters have $r_h/r_t \leq 0.03$.  Those characterised by $r_h/r_t \gtrsim 0.03$ (which appear only when $r_c=0.7$\,pc; case {\it (a)}) have a mass of order $100\,M_\odot$ and become therefore rapidly irrelevant to the cluster age distribution.  As an example, a cluster with a mass of $10^4\,M_\odot$ and a half-mass radius of 0.8\,pc orbiting the Milky Way halo at a galactocentric distance of 37\,kpc has $r_h/r_t \simeq 0.01$.  In a forthcoming paper, we will extend our model to stronger tidal fields, so as to assess the impact of the environment on the early star cluster history.

Once the bound fraction $F_b(t, \epsilon,\tau _{\rm GR}/\tau _{\rm cross}, r_h/r_t)$ has been estimated, the instantaneous (bound) star cluster mass corresponding to each star forming core is obtained via equation \ref{eq:minit}.  The corresponding temporal evolution of the mass in star clusters is then straightforwardly built.  We have simulated 18 different star cluster systems, combining the three core mass-radius relations described above with two mean SFEs ($\bar\epsilon=0.25$ and $0.40$) and three standard deviations ($\sigma _{\epsilon}=0.01, 0.04, 0.07$) for the SFE distribution function $P(\epsilon)$.  Results of these simulations are illustrated in Figs.~\ref{fig:cfr0.25} and \ref{fig:cfr0.40}.

Each panel shows the time evolution of the mass in dense core gas available to star formation per time unit (solid lines with asterisks), the star formation rate (SFR; dash-dotted lines with plain circles), and the resulting age distribution of the mass in star clusters ('SCs': dotted lines with open triangles, squares, circles for $\sigma _{\epsilon}=0.01, 0.04, 0.07$, respectively).  All mass counts are in unit of $1\,M_{\odot}.yr^{-1}$.  Because of model hypotheses, histories of the SFR and of the core formation rate are time-independent.  In contrast, the combined infant mortality and infant weight-loss of star clusters is clearly highlighted.  About 8\,Myr after gas expulsion, the star cluster age distribution starts decreasing as a result of unbound stars reaching cluster tidal limits.  At an age of $\simeq$ 30-50\,Myr, violent relaxation is over and the age distribution stabilizes (provided that a fraction of the initial gas-embedded cluster population survives).  That the age distribution eventually reaches an almost constant value is also due to our assumption of a weak tidal field for the simulated star cluster systems, which is appropriate for, e.g, the Small and  Large Magellanic Clouds~\citep[see][, respectively]{bou03,par08a}.  In dense environments, such as the Whirlpool galaxy M51 for which the dissolution time-scale -- due to secular evolution -- of a $10^4\,M_{\odot}$ is $\simeq 100$\,Myr~\citep[]{gie05}, the cluster age distribution will keep declining, albeit at a different rate, as a result of the sole cluster secular evolution (two-body relaxation and tidal interactions with the host galaxy).

Figs.~\ref{fig:cfr0.25} and \ref{fig:cfr0.40} show that the post-violent-relaxation cluster age distribution is a sensitive function of $P(\epsilon)=G(\bar\epsilon, \sigma _\epsilon)$.  (Note that the lower limit of the $y$-scale in Fig.~\ref{fig:cfr0.40} is much higher than in Fig.~\ref{fig:cfr0.25}). In case of instantaneous gas expulsion
($\tau _{\rm GR} << \tau _{\rm cross}$) and weak tidal field, the SFE threshold $\epsilon_{th}$ required to get a bound cluster at an age of $\gtrsim 50$\,Myr is $\epsilon_{th} \simeq 0.33$~\citep[Fig.~1 in][]{par07}.  Therefore, any $P(\epsilon)$ function leading to a negligible contribution of the $\epsilon > \epsilon_{th}$ regime (e.g. $P(\epsilon)=G(0.25, 0.01)$) leads to the disruption of the whole cluster population on a time-scale $\lesssim$ 30\,Myr.  In contrast, when $\bar\epsilon = 0.40$, the vast majority of the clusters survive, albeit depleted by about 65\,per cents of their mass, i.e. $F_b(t=50\,Myr, \epsilon =0.40) \simeq 0.35$~\citep[Fig.~1 in][]{par07}.  The decrease with time in the mass in star clusters is, in this case, an imprint of infant weight-loss only.  Generally speaking, when $\sigma$ is small, the bound fraction $F_b(t=50\,Myr, \epsilon )$ is an excellent proxy to the ratio between the age distributions of the masses in clusters and in stars for $t \gtrsim 50$\,Myr.  When $\bar\epsilon < \epsilon_{th}$, the final cluster-to-star mass ratio depends on $\sigma$ since it determines how large the contribution of the $\epsilon > \epsilon_{th}$ regime is (see also the discussion in the next section).

The assumption of either core mass-radius relation mostly impacts on the history of the mass in clusters when the SFE is low.  Considering the case of $P(\epsilon)=G(0.25, 0.04)$ (dotted curves with open squares in Fig.\ref{fig:cfr0.25}), the age distribution at $t \gtrsim 50$\,Myr is increasing when the core mass-radius relation gets shallower (from panel {\it c} to {\it a}; in the extreme case, when the core radius is constant).  The origin of this behaviour is to be sought in Fig.~\ref{fig:tt_mc} which depicts the relation between the gas expulsion time-scale $\tau _{\rm GR}/\tau _{\rm cross}$ and the core mass $m_c$ for the different core mass-radius relations used in our simulations.  The thick horizontal line at $\tau _{\rm GR}/\tau _{\rm cross} \simeq 1/3$ parts the regime of explosive gas expulsion from the adiabatic one.  Actually, inspection of Fig.~1 in \citet{par08b}, which shows $F_b(t=50\,Myr, \epsilon , \tau _{\rm GR}/\tau _{\rm cross})$ for a weak tidal field, illustrates that as long as $\tau _{\rm GR}/\tau _{\rm cross} \leq 1/3$,
the bound fraction is constant and characteristic of instantaneous gas removal.  Once the protocluster gas is expelled on a time-scale slower than $\tau _{\rm GR}/\tau _{\rm cross} \simeq 1/3$, the bound fraction is steadily increasing with the gas expulsion time-scale.  Fig.~\ref{fig:tt_mc} shows that when $r_c \propto m_c^{1/2}$, all gas cores undergo explosive gas expulsion.  In contrast, when $r_c$ is constant, cores more massive than $10^5\,M_{\odot}$ undergo adiabatic gas expulsion\footnote{It is worth keeping in mind that the quantitative results of Fig.~\ref{fig:tt_mc} depend on the normalisations we have adopted for each of the core mass-radius relations.  A core radius twice as large for any given core mass would make gas expulsion almost twice as fast (equation 6 in \citet{par08b})}.  The bound fraction, for any given age, is then higher than for instantaneous gas expulsion and is a sensitive function of $\tau _{\rm GR}/\tau _{\rm cross}$.  As a result, the stellar mass fraction remaining in bound clusters gets higher as the contribution of the adiabatic regime is growing.

A puzzling feature in the history of the cluster mass when $\bar\epsilon = 0.25$ is that it {\it increases} over the age range 50-100\,Myr for shallow core mass-radius relations (e.g. open triangles in panel {\it a} and open squares in panels {\it a} and {\it b} of Fig.~\ref{fig:cfr0.25}), while one would expect a stabilizing age distribution.  This is a consequence of the size-of-sample effect for the core mass spectrum, combined to the ability of high-mass cores to undergo adiabatic gas expulsion when $r_c \propto m_c^{1/3}$ or when $r_c$ is constant (Fig.~\ref{fig:tt_mc}).
Let us consider the oldest age bin for $\sigma = 0.01$ and $r_c = constant$ (rightmost open triangle in panel {\it a}).  It covers the age range $7.8 \leq {\log} (t) \leq 8.0$ and is contributed to by three clusters only, all arising from cores more massive than $5 \times 10^6\,M_{\odot}$.  Because of the assumed constant core radius, these very massive cores undergo slow gas expulsion and, therefore, retain a non-zero fraction of their stars (of order $\simeq 50$\,per cents for two of them) despite an SFE $\epsilon \simeq 0.25$.   
Less massive cores, owing to their shallower gravitational potential, result into disrupted clusters ($F_b = 0$).  At younger age ($7.4 < {\log} (t) < 7.6$), constant logarithmic age bins cover shorter linear age ranges, resulting into a smaller number of cores per bin, and a decreased probability of including very massive cores able to provide bound star clusters.

\section{Discussion}
\label{sec:disc}

The previous section has highlighted the steep relationship binding the mean local SFE $\bar\epsilon$ and the history of the mass in star clusters under the assumptions of a weak external tidal field and of a  constant gas-embedded cluster formation rate.  Table \ref{tab:clusterfrac} lists, for the 18 runs of Monte-Carlo simulations, the global (i.e. integrated over the age range 1-100\,Myr) cluster formation efficiency $\epsilon _{cl}$ and cluster-to-star mass ratio $M_{cl}/M_{st}$.  The cluster formation efficiency $\epsilon _{cl}$ is defined as the ratio between the total mass in stars bound to the clusters over the age range 1-100\,Myr and the total mass in core gas.  These ratios appear robust with respect to both the assumed core mass-radius relation and the standard deviation $\sigma _{\epsilon}$ of the SFE distribution function.
{\it Table \ref{tab:clusterfrac} demonstrates that, for this set of simulations, the cluster formation efficiency $\epsilon _{cl}$ and the cluster mass-to-star mass ratio $M_{cl}/M_{st}$ are sensitive tracers of the local mean SFE $\bar\epsilon$}.  Depending on the core mass-radius relation and SFE standard deviation, the ratio between the total mass in clusters and that in stars is 4 to 7 times higher when the mean SFE is $\bar\epsilon = 0.40$ than when $\bar\epsilon = 0.25$.  This tight relation is a direct consequence of residual star forming gas expulsion out of protoclusters.  For a population of clusters which have survived their violent relaxation and which are not yet affected by secular evolution significantly (or equivalently which are subject to a weak tidal field), the ratio between the total mass in clusters and the total mass in stars is simply given by:
\begin{equation}
\frac{M_{cl}}{M_{st}}(50 \lesssim t \lesssim 100\,Myr) \simeq \frac{\overline{F_{bound}} \times {\bar \epsilon} \times M_c}{\bar{\epsilon} \times M_c} \simeq \overline{ F_{bound}}\,,
\label{eq:fbound}
\end{equation}
where $M_c$ is the total mass in core gas, $\overline{F_{bound}}$ and $\bar{\epsilon}$ are the bound fraction $F_b(50 \lesssim t \lesssim 100\,Myr, \epsilon)$ and the SFE averaged over the whole population of clusters, respectively. The cluster mass fraction, as compared to the total mass in stars formed in the same star formation episode, is thus governed by the post-violent-relaxation bound fraction which is itself a steep function of the local SFE~\citep[see Fig.~1 of][for the case of a weak tidal field]{par07}\footnote{Note that $M_{cl}/M_{st}$ differs from 0 in Table~\ref{tab:clusterfrac} even for $\bar \epsilon =0.25$, $\sigma _{\epsilon}=0.01$ and explosive gas expulsion because the cluster-to-star mass ratio in this table is integrated over the age range 1-100\,Myr and, therefore, accounts for clusters still little affected by violent relaxation}.  
Measurements of the flux ratio from star clusters relative to field stars in young starbursts combined to improved models accounting, when required, for a significant external tidal field and a non-constant gas-embedded cluster formation rate thus constitute a promising way of probing the local SFE in active star forming environments, without resorting to gas mass estimates. 
Prerequisites include that the age range of the star cluster system includes the post-violent-relaxation age range, i.e. [50,100]\,Myr for a weak tidal field.  If the tidal field is stronger, cluster violent relaxation is faster since protocluster tidal radii are smaller~\citep[][]{bau07}.  

\begin{table}
\begin{center}
\caption{Ratio between the total mass in clusters and the total mass in stars ($M_{cl}/M_{st}$) and cluster formation efficiency $\epsilon _{cl}$ (i.e. ratio between the total mass in clusters and the total mass in core gas) for all simulations shown in Figs.~\ref{fig:cfr0.25} and \ref{fig:cfr0.40}.  All total masses are integrated over the age range 1-100\,Myr \label{tab:clusterfrac}.  A weak tidal field and a constant gas-embedded cluster formation rate are assumed.  Bracketted values correspond to cluster-to-star mass ratios for clusters brighter than $M_V = -7.3$, thereby emulating a detection limit} 
\begin{tabular}{c c c c c c c c c} \tableline 
                  &    &  \multicolumn{3}{c}{$\bar \epsilon = 0.25$} & & \multicolumn{3}{c}{$\bar \epsilon = 0.40$}   \\ \tableline
$\sigma _\epsilon$ & ~~~ & ~$0.01$~ & ~$0.04$~ & ~$0.07$~ & ~~~ & ~$0.01$~ & ~$0.04$~ & ~$0.07$~  \\ \tableline
\multicolumn{9}{c}{(a) $ r_c = constant $} \\ \hline

$M_{cl}/M_{st}$       &     &  $0.06$  &  $0.06$  &  $0.10$  &     &  $0.42$  &   $0.42$ &  $0.43$   \\
                &     &  $(0.04)$  &  $(0.04)$  &  $(0.07)$  &     &  $(0.27)$  &   $(0.27)$ &  $(0.27)$   \\
$\epsilon _{cl}$      &     &  $0.02$  &  $0.02$  &  $0.03$  &     &  $0.17$  &   $0.17$ &  $0.17$   \\ \tableline

\multicolumn{9}{c}{(b) $ r_c \propto m_c^{1/3} $} \\ \hline

$M_{cl}/M_{st}$       & ~~~ & ~$0.07$~ & ~$0.08$~ & ~$0.10$~ & ~~~ & ~$0.36$~ & ~$0.36$~ & ~$0.39$~  \\ 
    & ~~~ & ~$(0.06)$~ & ~$(0.06)$~ & ~$(0.07)$~ & ~~~ & ~$(0.23)$~ & ~$(0.22)$~ & ~$(0.25)$~  \\ 
$\epsilon _{cl}$      &     &  $0.02$  &  $0.02$  &  $0.03$  &     &  $0.15$  &  $0.14$  &  $0.16$   \\ \tableline

\multicolumn{9}{c}{(c) $ r_c \propto m_c^{1/2} $} \\ \hline
 
$M_{cl}/M_{st}$       & ~~~ & ~$0.08$~ & ~$0.08$~ & ~$0.10$~ & ~~~ & ~$0.36$~ & ~$0.36$~ & ~$0.40$~  \\
  & ~~~ & ~$(0.06)$~ & ~$(0.07)$~ & ~$(0.08)$~ & ~~~ & ~$(0.23)$~ & ~$(0.24)$~ & ~$(0.27)$~  \\
$\epsilon _{cl}$      &     &  $0.02$  &  $0.02$  &  $0.03$  &     &  $0.15$  &  $0.15$ &   $0.16$   \\ \tableline

\end{tabular}
\end{center}
\end{table}

Table \ref{tab:clusterfrac} also shows that cluster mass fractions as high as 40 per cents are not surprising: they are the signature of a local SFE only slightly in excess of the threshold $\epsilon _{th}$ required to form a bound star cluster.  If $\bar\epsilon \simeq 0.50$, the cluster mass fraction leaps to 70 per cents.  It is even higher in case of slow gas expulsion~\citep[Fig.~1 in][]{par07}, as is the case for small size massive cores.

\begin{figure}
\begin{center}
\epsscale{1.2} \plotone{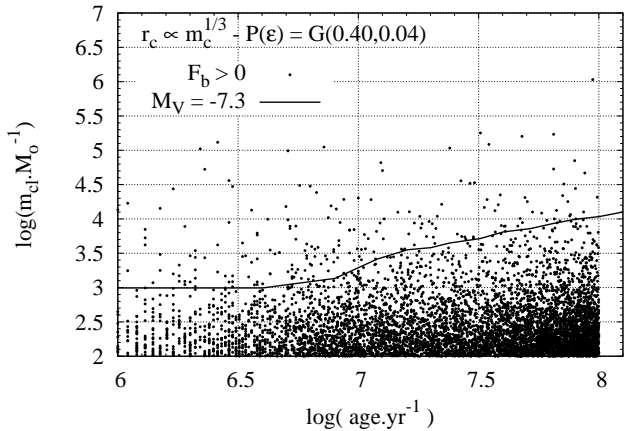}
\caption{Age-mass diagram of the fiducial star cluster system with $r_c \propto m_c^{1/3}$ and $P(\epsilon)=G(0.40, 0.04)$.  The solid line corresponds to a constant magnitude of $M_V^{\rm lim} = -7.3$.
A fraction only of the clusters arising from the one million cores simulated is shown}
\label{fig:detect}
\end{center} 
\end{figure}

However, it must be kept in mind that these cluster mass fractions are upper limits.  Firstly, at a given age, the presence of a strong tidal field, as may be expected in dense star forming environments, would make them lower~\citep[see Table 1 in ][]{bau07}.  Secondly, the census of star clusters in a distant galaxy is necessarily incomplete and the detection limit implies that the lower cluster mass range is missed.  We now investigate this second point, that is, how much the cluster-to-star mass ratio is affected by the presence of a detection limit.  Should it not be corrected for, the $M_{cl}/M_{st}$ ratio will be underestimated and an erroneously low local SFE estimate will be inferred.  We illustrate this issue with a 100 per cent-completeness limit at $M_V^{\rm lim} = -7.3$, which is comparable to what has been found by \citet{bas05} in their {\sl HST} study of the M51 star clusters, at a distance of 8.4\,Mpc.  All star cluster masses were converted into luminosities with the GALEV model~\citep{sch02} and an additional cluster-to-star mass ratio was computed, which takes into account clusters brighter than $M_V^{\rm lim} = -7.3$ only.  These are the bracketted values in Table \ref{tab:clusterfrac}. As an example, the age-mass diagram of the star cluster system with $r_c \propto m_c^{1/3}$ and $P(\epsilon)=G(0.40, 0.04)$, along with the adopted detection limit, is shown in Fig.~\ref{fig:detect}.  Although the vast majority of the clusters are fainter than $M_V^{\rm lim} = -7.3$, the $M_{cl}/M_{st}$ ratio is reduced by about one third only.  This is due to the choice of a power-law of spectral index $-2$ for the core mass spectrum.  While clusters with a mass in the range of, say, $100\,M_\odot < m_{cl} < 3000\,M_\odot$ (where the upper limit is chosen as a rough averaged estimate of the cluster mass at the detection limit) represent 97 per cents of the cluster number, they contribute by 30 per cents only to the total mass of the star cluster system.  Their non-detection therefore does not severely impact on the $M_{cl}/M_{st}$ ratio.  The effect will of course be more prominent as the detection limit gets brighter.  Knowledge of the detection limit and of the age range covered by the star cluster system, combined to detailed simulations such as those carried out in this paper, are necessary to convert the observed $M_{cl}/M_{st}$ ratio into the actual one.  We also note that in distant galaxies, observed cluster-to-star mass ratios may be affected by confusion effects, i.e. groups of otherwise unbound stars being misinterpreted as genuine bound star clusters, which results in an overestimated total mass in clusters.  How these two effects -- detection limit and confusion -- affect the cluster-to-star mass/flux ratios derived by \citet{meu95} and \citet{deg03} remains to be seen. \\

In a recent study, \citet{bas08} argues that the formation of bound clusters -- those which survive violent relaxation -- represents, on the average, 8\,\% of the total star formation of a galaxy, irrespective of its SFR.  {\it If} the star clusters displayed in the relation between the present SFR of galaxies and the magnitude of their brightest young clusters \citep[][his Fig.~1]{bas08} have actually reached the endpoint of their violent relaxation, then we can use this result to estimate the corresponding local SFE.  If all stars form in gas-embedded clusters, the 8\,\% ratio then corresponds to our bound fraction $F_{bound}$, which gives a back-of-the-envelope estimate for the local star formation efficiency of $\epsilon \simeq 35\,\%$ \citep[assuming explosive gas expulsion and weak tidal field; Fig.~1 in][]{par07}, regardless of the host galaxy.  According to \citet[][]{ll03}, however, gas-embedded clusters account for 70-90\,\% of all stars formed in Galactic disk GMCs.  If we now assume that 30\,\% of all star formation in a galaxy takes place in the field or within loose stellar associations,
the bound fraction, i.e. the ratio between the total mass in bound star clusters and that in gas-embbeded star clusters, becomes $F_{bound} \simeq 8\,\%/0.7 \simeq$ 11\,\% which, by virtue of the sharp dependence of $F_{bound}$ on $\epsilon$ (especially in the $\epsilon \gtrsim \epsilon _{th}$ regime), leads to an only  marginally higher mean local SFE.

The combination of the result of \citet{bas08} with our model provides therefore a hint for a local SFE which does not vary significantly among galaxies of the local Universe.  This is consistent with fig.~2a in \citet[][]{gao04} which shows that the ratio between the total far-IR luminosity $L_{IR}$, a tracer of the SFR, and the HCN line luminosity $L_{HCN}$, a tracer of the dense star forming gas, is roughly constant (although with much scatter) over 4 orders of magnitude in SFR, from almost passive spirals to the strongest ULIRGs.  It is worth noting that the local SFE estimate quoted above is only slightly higher than the SFE threshold $\epsilon _{th}$ for bound cluster formation in the regime of explosive gas expulsion. Therefore, even slight variations around the average value $\epsilon \simeq 0.35$ are conducive to significant variations in the bound cluster-to-star mass ratio (say, from 0\,\% to 35\,\% for an SFE range of $\epsilon \lesssim $30\,\% to $\epsilon \simeq $40\,\%).

Merging galaxies have high SFRs, the origin of which is in the large fraction of molecular gas ``trapped" in the dense phase of star forming cores, rather than in the more diffuse phase of GMCs.  For instance, ULIRGs have conspicuously high mass ratio of dense star forming gas compared to their total molecular gas, as traced by the HCN line luminosity to CO luminosity ratio $L_{HCN}/L_{CO}$~\citep{Solomon+97,san96}.  This is the very reason of their 1-to-2 orders of magnitude higher SFR compared to spiral galaxies~\citep[][their fig.~4]{gao04}.
Because of the observed correlation between the SFR and the visual absolute magnitude of the brightest young cluster \citep[][]{lar02,wkl04}, increased SFRs in turn gives rise to the formation of more massive clusters, which are more likely to survive several Gyr of secular evolution.  This is why major star formation episodes of a galaxy are well-traced by peaks in its cluster age distribution.

\section{Model caveats}
\label{sec:cav}

Table \ref{tab:clusterfrac} shows, for the models explored in section 2 (i.e. weak tidal field and constant gas-embedded cluster formation rate), the ratio between the total mass in clusters and the total mass in stars formed in gas-embedded clusters over the age range 1 to 100\,Myr (i.e. this ratio accounts for clusters of which the stellar content is still unaffected by violent relaxation).  It appears that these cluster-to-star mass ratios do not sensitively depend on either the standard deviation $\sigma _\epsilon$ of the SFE distribution function or the core mass-radius relation.  In contrast, the ratio between the total mass in bound clusters and that in stars (i.e. the cluster-to-star mass ratio integrated over the age range 50-100\,Myr) {\it does} depend on the fine details and underlying model assumptions, as can be seen from Fig.~\ref{fig:cfr0.25} and Fig.~\ref{fig:cfr0.40}.  Because an increased bound cluster-to-star mass ratio can be caused by either an increased mean local SFE $\bar \epsilon$, an increased SFE distribution function standard deviation $\sigma _\epsilon$, or a shallower core mass-radius relation $m_c-r_c$, the model is prone to degeneracies.  
As for the mass-radius relation, the influence of gas expulsion on the slope of the cluster mass-radius relation as compared to that of their progenitor cores must be studied.  In a forthcoming paper, we will apply our model to the data of Fig.~1 of \citet{bas08} so as to constrain more tightly the distribution function of the local SFE.  That is, we will investigate in greater detail what is the range in local SFE permitted by the data scatter around the linear relation between the logarithm of the present SFR and the visual absolute magnitude of the brightest young cluster in galaxies.

We also remind the reader that our present results are valid only in the case of a weak tidal field.
Strong tidal fields both decrease the ability of gas-embedded clusters to survive infant weight-loss and shorten the violent relaxation phase, in essence because a smaller cluster tidal radius enhances cluster loss of stars \citep{bau07}.  One way of assessing the tidal field strength in a given (region of a) galaxy is to estimate the cluster dissolution time-scale due to secular evolution, based on the age and mass distributions of clusters older than 50\,Myr \citep[see e.g.][]{bou03,gie05,par08a}.  One can then match this time-scale to the external tidal field via equation 7 of \citet[][]{bau03}.  

Finally, our simulations assume that the mass of dense core gas available to star formation per unit of time and the SFR remain constant with time.  If a galaxy is entering a ULIRG phase, the amount of dense star forming gas will increase as time goes by, boosting the SFR and CFR at young ages.  Should this effect not be corrected for, one could -- wrongly -- infer that the decrease in the cluster age distribution towards old ages is entirely due to cluster destruction and, therefore, underestimate the local SFE.

\section{Conclusions}

Building on the $N$-body simulation grid recently obtained by \citet{bau07}, we have performed detailed simulations of the early (i.e. up to an age of 100\,Myr) evolution of star cluster systems.  In this first contribution, we have adopted the following assumptions: star forming cores with a power-law mass spectrum of slope $-2$, constant gas-embedded cluster formation rate, Gaussian distribution function of the local SFE and weak external tidal field.  Besides, star formation is assumed to mostly take place in gas-embedded clusters.  Under these assumptions, we have highlighted the strong impact of the local SFE on the temporal evolution of the mass in clusters and on the ratio between the total mass in clusters and the total mass in stars.  Within less than 10\,Myr after gas expulsion, the age distribution of the mass in clusters decreases as a result of cluster violent relaxation and the amplitude of the decline depends strongly on the adopted SFE distribution function.  This is because the fraction of stars remaining bound to a cluster after its violent relaxation is sharply increasing with the SFE $\epsilon$ once 
$\epsilon > 0.33$~\citep[Fig.~1 in][]{par07}.  The ratio $M_{cl}/M_{st}$ between the total mass in stars still bound to the clusters and the total mass in stars formed in gas-embedded clusters over the age range of $1-100$\,Myr constitutes a sensitive probe of the local SFE which characterized their progenitor cores.  Besides, this ratio is only weakly affected by the choice of a core mass-radius relation.
We caution however that {\it (i)} observationally derived cluster-to-star mass ratios account for detected clusters only; {\it (ii)} ratios displayed in Table \ref{tab:clusterfrac} will be different for other tidal field strengths and non-constant gas-embedded cluster formation history.  
Measuring the flux ratio given off by star clusters in a young starburst and converting it in a cluster-to-star mass ratio therefore constitutes a promising way of inferring the local SFE, provided that the observed $M_{cl}/M_{st}$ is turned into the actual one and that time-varying gas-embedded cluster formation rate and stronger tidal fields are -- if necessary -- accounted for via generalized versions of the present model.  

Another potential application of our model is to trace the evolution of the local SFE over cosmological lookback times.  This can be done, for galaxies close enough so that their field stars can be resolved and age-dated, by comparing the history of the bound cluster formation rate (i.e. the age distribution of the cluster mass at the end of violent relaxation, obtained by correcting the observed age distribution for secular evolutionary mass losses and fading effects, see e.g. \citet{par08a}) on the one hand and the history of the SFR on the other hand.  This bound cluster-to-star mass ratio is our bound fraction $F_{bound}$ (see equation \ref{eq:fbound}), equivalent to model outputs over the age range 50-100\,Myr.  This application, however, requires tighter constraints on the slope of the core mass-radius relation, the width of the local SFE distribution and an estimate of the external tidal field strength.

Combined with the recent finding of \citet{bas08} that, on the average, galaxies form 8\,\% of their stars in bound star clusters, the relation between the bound fraction $F_{bound}$ and the local SFE $\epsilon$ constitutes a hint towards an almost universal local SFE in the present-day Universe.  In a forthcoming paper, we will apply our model to the observed relation between the present-day SFR of galaxies and the absolute visual magnitude of their brightest young clusters so as to shed better constraints on the present-day local SFE distribution function.



\acknowledgments
GP acknowledges support from the Belgian Science Policy Office in the form of a Return Grant 
and from the Alexander von Humboldt Foundation in the form of a Research Fellowship.  
GP acknowledges support and hospitality of the Cambridge Institute of Astronomy and the 
Hertfordshire Centre of Astrophysics where this work was started.
GP and UF are grateful for research support and hospitality at the 
International Space Science Institute in Bern (Switzerland), 
as part of an International Team Programme.  We thank an anonymous referee for a critical
and constructive report.  GP thanks Holger Baumgardt for useful comments about his $N$-body
simulation model grid and Nate Bastian and Soeren Larsen for useful discussions
at the ''From Taurus to the Antennae`` conference.







\end{document}